\documentclass[a4paper,10pt,oneside,onecolumn,preprint]{elsarticle}
\usepackage {graphicx}

\usepackage{color}

\begin {document}

\title {The influence of defects on magnetic properties of fcc-Pu}

\author {A.~O.~Shorikov}
\author {M.~A.~Korotin}
\author {V.~I.~Anisimov}

\address {Institute of Metal Physics, Ural Division of Russian Academy
of Sciences,  18, S. Kovalevskaya str., 620041 Ekaterinburg GSP-170, Russia}

%\addres {Institute of Metal Physics, Ural Division of Russian Academy
%of Sciences,  18, S. Kovalevskaya str., 620041 Ekaterinburg GSP-170, Russia}

%\address {Institute of Metal Physics, Ural Division of Russian Academy
%of Sciences,  18, S. Kovalevskaya str., 620041 Ekaterinburg GSP-170, Russia}

\author {V.~V.~Dremov}%\corref{cor1}\fnref{fn1}}
\ead{v.v.dryomov@vniitf.ru}
\author {Ph.~A.~Sapozhnikov}

%\address {Russian Federal Nuclear Center - Institute of Technical
%Physics, 13, Vasiliev str., Snezhinsk, 456770 Chelyabinsk region, Russia}

\address {Russian Federal Nuclear Center - Institute of Technical
Physics, 13, Vasiliev str., Snezhinsk, 456770 Chelyabinsk region, Russia}
%\fntext[fn1]{v.v.dryomov@vniitf.ru }
\begin {abstract}

The influence of vacancies and interstitial atoms on magnetism in Pu
has been considered in frames of the Density Functional Theory
(DFT). The relaxation of crystal structure arising due to different
types of defects was calculated using the molecular dynamic method
with modified embedded atom model (MEAM). The LDA+U+SO (Local
Density Approximation with explicit inclusion of Coulomb and
spin-orbital interactions) method in matrix invariant form was
applied to describe correlation effects in Pu with these types of
defects. The calculations show that both vacancies and interstitials
give rise to local moments in $f$-shell of Pu in good agreement with
experimental data for annealed Pu. Magnetism appears due to
destroying of delicate balance between spin-orbital and exchange
interactions.

\end {abstract}

\maketitle

\section {Introduction}

Band structure calculations of $\delta$-Pu predict the static
magnetic order of $f$-electrons with the values of full magnetic
moment 0.25--5~$\mu_B$ with substantial impact of spin moment
~\cite{Savrasov00,Lieser74,Bouchet00,Soderlind02}. These
results contradict to experimental measurements of magnetic
properties of non-aged Pu without impurities. These data indicate
the absence of any ordered or disordered, static or dynamic magnetic
moments in Pu at low temperatures~\cite {Lashley05, Fluss06}.

Resent development of calculation methods allows to describe
correctly the ground state of pure Pu in $\delta$- and model $\alpha$-phase~\cite {Shick05, shorikov05}.
It has been shown in Ref.~\cite {shorikov05} that the delicate balance between
spin-orbit (SO) and exchange interactions determines the nonmagnetic
ground state in pure Pu. These interactions have the magnitude close
to each other in actinides and its compounds and the balance could
be easily broken by crystal field of legands. 
Also, P. S\"oderlind~\cite{Soderlind08}
confirs the important role of SO and orbital polarization in formation 
of nonmagnetic ground state of plutonium in frames of model DFT calculation.
Impurities like Al and
Ga that are used to stabilize the fcc-phase of Pu act in the same
way. Several groups report the presence of ordered magnetic moment
in aged Pu-Al and Pu-Ga alloys~\cite {Heffner05, Verkhovskii05,
Verkhovskii07}. The magnitude of moments is small,
$\leq$10$^{-3}$~$\mu_B$ (Ref.~\cite {Heffner05}) --
0.15~$\mu_B$ (Ref.~\cite {Verkhovskii05, Verkhovskii07}), and
these moments could arise due to distortion of crystal structure
near interstitial Pu atoms and vacancies.

Substantial drawback of the Local (Spin) Density Approximation (L(S)DA) is the
underestimation of orbital moment~\cite {Singh, Chen}. This feature
leads to fail of the DFT in description of $4f$- and $5f$-metals,
since orbital moment in them can overcome spin one. Taking into
account the Coulomb repulsion U and SO interactions
in full matrix rotation invariant form in the LDA+U+SO method could
improve the results. Achievement of this method is that one
shouldn't set the exact magnetic order at the start of iterations.
Both magnitude and direction of magnetic moment are calculated for
each atom. Magnetic order and direction of ``easy axis'' are the
result of self-consistent interaction procedure. 

In the LDA+U method\cite{LDA+U} the energy functional $E_{LDA+U}$ depends, 
in addition to the charge density $\rho (\mathbf{r)}$, on the
occupation matrix $n_{mm^{\prime }}^{s s^{\prime }}$ for particular 
orbital for which correlation effects are taken into account (in our 
case it is 5$f$ plutonium orbitals). The LDA+U method
in general nondiagonal in spin variables form was defined
in Ref.~\cite{Solovyev98}
\begin{equation}
E_{LDA+U}[\rho (\mathbf{r}),\{n\}]=E_{LDA}[\rho (\mathbf{r)}%
]+E_U[\{n\}]-E_{dc}[\{n\}]  \label{U1}
\end{equation}
where $\rho (\mathbf{r})$ is the charge density, $E_{LDA}[\rho
(\mathbf{r})]$ is the standard LDA functional. The occupation
matrix is defined as
\begin{equation}
n_{mm^{\prime }}^{s s^{\prime }}=-\frac 1\pi
\int^{E_F}ImG_{mm^{\prime }}^{s s^{\prime }}(E)dE \label{Occ}
\end{equation}
where $G_{mm^{^{\prime }}}^{s s^{\prime }}(E)=\langle ms \mid
(E-\widehat{H}_{LDA+U})^{-1}\mid m^{\prime }s^{\prime
}\rangle $ are the elements of the Green function matrix in local
orbital basis set ($m$ -- magnetic quantum number, and $s$ -- spin 
index for correlated orbital). In the present work this basis set 
was formed of LMT-orbitals from the tight binding LMTO method based 
on the atomic sphere approximation (TB-LMTO-ASA).~\cite{LMTO} 
In Eq. (\ref{U1}) Coulomb interaction energy $E_U[\{n\}]$ term is a
function of occupation matrix $n_{mm^{\prime }}^{s s^{\prime }}$
\begin{equation}
\begin{array}{c}
E_U[\{n\}]=\frac 12\sum_{\{m\},ss^{\prime }}\{\langle
m,m^{\prime \prime }\mid V_{ee}\mid m^{\prime },m^{\prime \prime \prime
}\rangle n_{mm^{\prime }}^{ss}n_{m^{\prime \prime }m^{\prime
\prime \prime }}^{s^{\prime }s^{\prime }} \\
\\
-\langle m,m^{\prime \prime }\mid V_{ee}\mid m^{\prime \prime
\prime },m^{\prime }\rangle n_{mm^{\prime }}^{ss^{\prime
}}n_{m^{\prime \prime }m^{\prime \prime \prime }}^{s^{\prime }s
}\}
\end{array}
\label{upart}
\end{equation}
where $V_{ee}$ is the screened Coulomb interaction between the correlated 
electrons. Finally, the last term in Eq. (\ref{U1}) correcting for
double counting is a function of the total number of electrons in
the spirit of LDA which is a functional of total charge density
\begin{equation}
E_{dc}[\{n\}]=\frac {1}{2}UN(N-1)-\frac {1}{4}J_HN(N-2) 
\label{U3}
\end{equation}
where $N$ = $Tr(n_{mm^{\prime }}^{ss^{\prime }})$ is a total number of
electrons in the particular shell. $U$ and $J_H$ are screened Coulomb and
Hund exchange parameters which could be determined in the constrain
LDA calculations.~\cite{Gunnarsson89,Anisimov91a} 
%in this paper is everywhere denoted as $J_H$ not to be confused
%with total moment operator $J$. 
The screened Coulomb interaction
matrix elements $\langle m,m^{\prime \prime }\mid V_{ee}\mid
m^{\prime },m^{\prime \prime \prime }\rangle$ could be expressed
via parameters $U$ and $J_H$ (see Ref.~\cite{LDA+U}).

The functional Eq. (\ref{U1}) defines the effective single-particle
Hamiltonian with an orbital dependent potential added to the
usual LDA potential
\begin{equation}
\widehat{H}_{LDA+U}=\widehat{H}_{LDA}+\sum_{ms,m^{\prime
}s^{\prime }}\mid ms \rangle V_{mm^{\prime }}^{ss^{\prime
}}\langle m^{\prime }s^{\prime }|,\label{hamilt}
\end{equation}
\begin{equation}
\begin{array}{c}
V_{mm^{\prime }}^{ss^{\prime }}=\delta _{ss^{\prime
}}\sum_{m^{\prime \prime},m^{\prime \prime \prime}}\{\langle m,m^{\prime \prime }\mid V_{ee}\mid m^{\prime
},m^{\prime \prime \prime }\rangle n_{m^{\prime \prime }m^{\prime \prime
\prime }}^{-s,-s }+ \\
\\
(\langle m,m^{\prime \prime }\mid V_{ee}\mid m^{\prime },m^{\prime \prime \prime }\rangle
-\langle m,m^{\prime \prime }\mid V_{ee}\mid m^{\prime \prime \prime },
m^{\prime }\rangle )n_{m^{\prime \prime }m^{\prime \prime \prime }}^{ss}\}- \\
\\
\left( 1-\delta _{ss^{\prime }}\right) \sum_{m^{\prime \prime},m^{\prime \prime \prime}}\langle
m,m^{\prime \prime }\mid V_{ee}\mid m^{\prime \prime \prime },m^{\prime
}\rangle n_{m^{\prime \prime }m^{\prime \prime \prime }}^{s^{\prime
}s} \\
\\
-U(N-\frac{1}{2})+\frac{1}{2}J_H(N-1).
\end{array}
\label{Pot}
\end{equation}

In this paper we used method LDA+U+SO which
comprises non-diagonal in spin variables LDA+U Hamiltonian 
Eq. (\ref{Pot}) with spin-orbit (SO) coupling term 
\begin{equation}
\begin{array}{c}
\widehat{H}_{LDA+U+SO}=\widehat{H}_{LDA+U}+\widehat{H}_{SO},\\
\\ 
\widehat{H}_{SO}=\lambda \cdot  \mathbf L \cdot  \mathbf S
\label{eq:LS}
\end{array}
\end{equation}
where $\lambda$ is a parameter of spin-orbit coupling.
In $LS$ basis SO coupling matrix has
diagonal $(H_{SO})^{s,s}_{m^{\prime},m}$ as well as off-diagonal in spin variables 
$(H_{SO})^{\uparrow,\downarrow}_{m^{\prime},m}$ and 
$(H_{SO})^{\downarrow,\uparrow}_{m^{\prime},m}$
non-zero matrix elements (in complex spherical harmonics)\cite{LandauSO}
\begin{equation}
\begin{array}{c}
(H_{SO})^{\uparrow,\downarrow}_{m^{\prime},m} =
\frac{\lambda}{2}
\sqrt{(l+m)(l-m+1)}(\delta_{m^\prime,m-1}),\\
\\
(H_{SO})^{\downarrow,\uparrow}_{m^{\prime},m} =
\frac{\lambda}{2}\sqrt{(l+m)(l-m+1)}(\delta_{m^\prime-1,m}),\\
\\
(H_{SO})^{s,s}_{m^{\prime},m} = \lambda m s \delta_{m^\prime,m}
\end{array}
\label{SO-nondiag}
\end{equation}
where $lm$ -- orbital quantum numbers, spin index $s$ = +1/2, --1/2. 
The pecularities of LDA+$U$+SO method and its implemenataion to the problem of 
pure Pu an several plutonium compounds were described in detales in Ref.~\cite {shorikov05}.

In the present work four different fcc-Pu supercells were investigated.
Namely: one interstitial (IS) Pu atom in 32-atoms supercell, vacancy
in 8-atoms supersell, and two 32-atoms supercells with both IS and
vacancy at minimal and large distances. Due to the presence of
defects the perfect fcc structure was to be distorted and therefore
the relaxation of crystall structure for all supercells under
investigation should  to be taken into account. Since LMTO method doesn't 
make possible to perform 
structure relaxation correctly we used  Classical
Molecular Dynamics (CMD) with the Modified Embedded Atom Model
(MEAM) by Baskes~\cite {Baskes1, Baskes2, Baskes3} as interatomic potential. The MEAM is the many-body
potential, i.e. interaction between a pair of atoms depends on the
local structure (on positions of their common neighbors). The
parametrization of the MEAM for pure plutonium and plutonium-gallium
alloys was given in Ref.~\cite {Baskes1} and presently the
potential is widely used in CMD simulations of plutonium properties
and processes in Pu caused by self-irradiation (Refs.~\cite
{Baskes1, Baskes2, Baskes3, Dremov1, Dremov2}).

\begin {table}
\caption {Magnetic properties calculated for 32-atoms supercell and
interstitial (IS) Pu atom. First column -- the labels of nonequivalent Pu
atoms. Second column -- distance between IS and Pu ion (\AA).
Next four columns: number of equivalent Pu atoms in subclasses
(n$_{atoms}$), calculated values for spin ($S$), orbital ($L$), and total
($J$) moments. Last four columns contain partial contributions of $f^6$
configurations and $jj$-type of coupling for 5$f$ shell of Pu ion,
effective magnetic moment (see text for explanations) and total number of
$f$-electons.}
%\vspace {0.5cm}
\centering
\begin {tabular}{c|c|c|c|c|c|c|c|c|c}
\hline
&D, \AA & n$_{atoms}$ & $S$ & $L$ & $J$& $f^6$,\% & $jj$,\% & $\mu_{eff}$ & n$_f$ \\
\hline
IS &  & 1 & .028 & .057 & .03 & 98.9 & 99.2 & .146 & 6.06\\
\hline
   && 4 & .260 & .341 & .08 & 96.7 & 91.7 & .234 &\\
\raisebox {2ex}[1pt][1pt]{Pu1} & \raisebox {2ex}[1pt][1pt]{2.79}
   & 2 & .065 & .059 & .01 & 99.8 & 97.8 & .061 &
\raisebox {2ex}[1pt][1pt]{5.69}\\
\hline
   && 4 & .216 & .310 & .09 & 96.3 & 93.2 & .256 &\\
\raisebox {2ex}[1pt][1pt]{Pu2} & \raisebox {2ex}[1pt][1pt]{4.03}
   & 4 & .197 & .277 & .08 & 96.8 & 93.8 & .238 &
\raisebox {2ex}[1pt][1pt]{5.79}\\
\hline
   && 4 & .506 & .633 & .13 & 94.9 & 83.5 & .271 &\\
\raisebox {2ex}[1pt][1pt]{Pu3} & \raisebox {2ex}[1pt][1pt]{5.22}
   & 8 & .380 & .472 & .09 & 96.3 & 87.7 & .236 &
\raisebox {2ex}[1pt][1pt]{5.77}\\
\hline
   && 4 & .750 & 1.111 & .36 & 85.6 & 75.7 & .456 &\\
\raisebox {2ex}[1pt][1pt]{Pu4} & \raisebox {2ex}[1pt][1pt]{6.96}
   & 2 & .471 & .574 & .10 & 95.9 & 84.6 & .246 &
\raisebox {2ex}[1pt][1pt]{5.73}\\
\hline
\end {tabular}
\label {res_IS_large}
%\vspace {0.5cm}
\end {table}

Adding of IS or vacancy into initial superecell made Pu atoms inequivalent.
That is why the different types of atoms in Tables below have additional numbers (e.g.,
Pu1 etc). The relaxation of crystal structure lowers the symmetry again,
and the new Pu classes have been divided on sub-classes (see Tab.~\ref
{res_IS_large}).  All calculations of the electronic structure and magnetic
properties were made using the Tight-Binding Linear Muffin-Tin Orbitals
method with Atomic Sphere Approximation (TB-LMTO-ASA computation
scheme). 
%Only Pu $f$-states have been
%considered as magnetic ones in our calculations. The rest states were
%treated in frames of the LDA approximation. 
In the LDA+U calculation scheme
the values of direct Coulomb ($U$) and Hund's exchange ($J_H$)  parameters should
be determined as the first step of calculation procedure. It can be done in
\textit {ab initio} way via the constrained LDA calculations~\cite
{Anisimov91a, Gunnarsson89}. In our calculations the Hund exchange
parameter $J_H$ was found to be $J_H$ = 0.48~eV. The value of Coulomb
parameter $U$ was set to 2.5~eV since this value provides the correct
equilibrium volume of $\delta$-Pu (see Ref.~\cite {shorikov05} for
the details).

\section {Interstitial Plutonium atom in 32-atoms supercell}
\label {IS_large}

At first 32-atoms supercell of fcc-Pu with one additional Pu atom has been considered. 
The supercell has
three coordination spheres around defect that is sufficient to describe 
relaxation of position of neighbor atoms.  Classical 
molecular dynamic method was applied of describe distortion of crystal structure. New positions of Pu
atoms in supercell were used in further calculation of electronic structure.  
Adding of one additional Pu atom lowers the symmetry of the cell. Four new inequivalent classes of
plutonium belonging to four different coordination spheres around IS arouse. 
Moreover  Pu atoms within each new class  become inequivalent due to different local
neighborhood.
% after relaxation of crystal structure. 
To take in to account this lowering of symmetry no
symmetrisation was applied in our electronic structure calculation. Since no additional symmetry
conditions was imposed on electronic subsystem magnitude of local moments on Pu sites and their
directions can be arbitrary and correspond to the minimum of total energy.

The LDA+U+SO calculations for metallic Pu in $\delta$ phase gave a
nonmagnetic ground state with zero values of spin $S$, orbital $L$, and
total $J$ moments~\cite {Shick05, shorikov05}.  Our calculation made for
32-atoms supercell with one IS shows that small local magnetic moments
develop on Pu sites. The magnitude of local moment depends on the distance
between center of distortion (IS) and corresponding Pu site. We argue that
this appears because of the break of balance between SO and exchange
interactions due to the relaxation of crystal structure. The results are
presented in Tab.~\ref {res_IS_large}. The partial contributions of $f^6$
configuration and $jj$-type of coupling to the final state could be
calculated in the following way. Total moment value is the same in both
coupling schemes ($jj$ or $LS$): $J$ = 0 for $f^6$ and $J$ = 5/2 for $f^5$.
If there is a mixed state $(1-x)\cdot f^6+x\cdot f^5$ then $x$ can be
defined as $x$ = $J$/2.5. Spin $S$ and orbital $L$ moment values for $f^6$
configuration are equal to zero in $jj$ coupling scheme and $S$ = 3, $L$ =
3 in $LS$ coupling scheme. For $f^5$ configuration they are $S$ = 5/14
$\approx$ 0.36, $L$ = 20/7 $\approx$ 2.86 in $jj$ coupling scheme and $S$ =
5/2, $L$ = 5 in $LS$ coupling scheme. One can define a mixed coupling
scheme with a contribution of $jj$ coupling equal to $y$ and of $LS$
coupling to $(1-y)$, correspondingly. In final state the calculated values
of orbital and spin moments will be:
\begin {equation}
L=x\cdot (2.86\cdot y +5\cdot (1-y))+(1-x)\cdot (0\cdot y +3\cdot (1-y)),
\label {LS-model1}
\end {equation}
\begin {equation}
S=x\cdot (0.36\cdot y +2.5\cdot (1-y))+(1-x)\cdot (0\cdot y +3\cdot (1-y).
\label {LS-model2}
\end {equation}

These formulas allow to determine the value of coefficient $y$. An
effective paramagnetic moment obtained from susceptibility measurements
using Curie-Weiss law can be calculated as
\begin {equation}
\mu_{eff}=g\cdot\sqrt{J\cdot(J+1)}\cdot\mu_B.
\label {m-eff}
\end {equation}

The problem is to define Lande $g$-factor which can be calculated for pure
$f^5$ and $f^6$ configurations in $LS$ or $jj$ coupling schemes. As for
$f^6$ configuration total moment $J$ = 0, one needs to calculate $g$-factor
for $f^5$ configuration only. For ground state of $f^5$ configuration in
$jj$ coupling scheme Lande factor is $g_{jj}$ = 6/7 $\approx$ 0.86. In $LS$
coupling scheme its value is $g_{LS}$ = 2/7 $\approx$ 0.29. As the latter
value is nearly three times larger than the former, $g_{jj}$ and $g_{LS}$
can give only upper and lower limits of $g$-factor for the case of
intermediate coupling. We could calculate weighted value of effective
moment using relative weights of $LS$- and $jj$-couplings obtained from
Eqs.~\ref {LS-model1},~\ref {LS-model2},~\ref {m-eff}.

\begin {figure}
\centerline {\includegraphics [clip=true,width=.85\columnwidth]
{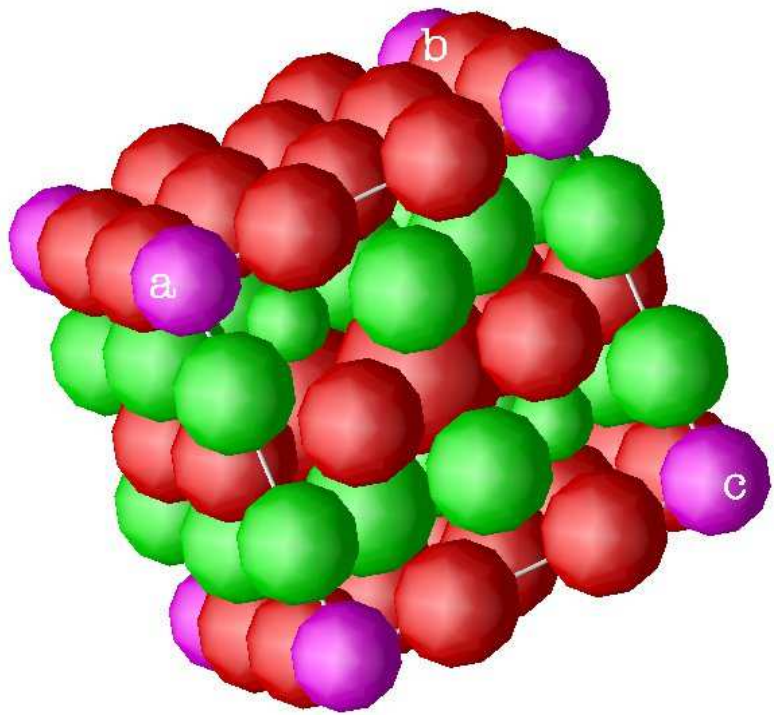} }
\caption {(Color online). Non-collinear order obtained for 32-atoms
supercell with IS. Purple spheres denote IS. Green and red spheres are Pu
atoms with opposite directed total moments. The radii of spheres are
proportional to the magnitude of correspondent magnetic moments.}
\label {afm-a.IS}
\end {figure}

The magnetic order of Pu ions is set arbitrary at the beginning of
iteration process. Final directions of local moments are calculated
in accordance with minimum of total energy at the end of
self-consistence loop. Since the long range order is set up in used
calculation method, some ferrimagnetic order arouse as an artifact.
This order resembles the antiferromagnetic (AFM) one of A-type. In
Fig.~\ref {afm-a.IS} resulting directions of total moment are shown
with red and green colors. Pu atoms positioned in the first
coordination sphere to IS as well as IS itself have the smallest
magnetic moments. They don't differ significantly from pure
$\delta$-Pu that is nonmagnetic. The values of local moments grow up
with increasing distance between IS and corresponding site. The
largest total moment developed for Pu4 ion positioned in the center
of supercell (large red sphere in Fig.~\ref {afm-a.IS}). This ion
has the largest distance to IS. The average value of the effective
moment in the case of IS Pu atom in 32-atoms supercell is
$\mu_{eff}\sim$0.26~$\mu_B$. Numbers of $f$-electrons (see
Tab.~\ref{res_IS_large} last column) differ from those calculated
for not distorted fcc-Pu (which has 5.74 $f$-electrons) but not
significantly except the case of interstitial atom. The later has
the largest occupation number in all considered structures (see
tables below). 
The large number of $f$-electrons (close to 6) obtained in  
the present calculation disagrees with previous experimental and 
theoretical estimations
that give 5.1-5.2 electrons\cite{Tobin08}. Such difference between 
theoretical results originates 
from dissimiliar band structure calculation methods. Since TB-LMTO-ASA scheme 
use artifically large overlaping atomic spheres these numbers should be 
considered
only to compare differnt clases of plutonium atoms with each other. 
We have verified our results and run several
calculations with differnet radii of Pu atoms filling of empty space in 
the primitive cell
with empty spheres (pseudo-atom without core states). Distorted supercell
always
becomes magnetic with the same order. The magnitude of local magnetic moments 
depends
slightly on atomic radius. It rises up with increasing of the later.       
For the sake of simplicity we have choosen the same radii 3.41 a.u. for all Pu 
atoms to be able to compare  
their magnetic moments. Artifical overlapping of atomic spheres in all 
considered supercells never overcomes 13\% 
which is critical TB-LMTO-ASA value. 

\section {Vacancy in 8-atoms supercell}
\label {ES_small}

\begin {table}
\caption {Magnetic properties of Pu ions calculated for 8-atoms supercell
with one vacancy. Second column -- distance between vacancy and Pu ion (\AA).
See also the caption of Tab.~\ref {res_IS_large}.}
%\vspace {0.5cm}
\centering
\begin {tabular}{c|c|c|c|c|c|c|c|c|c}
\hline
& D, \AA &n$_{atoms}$& $S$ & $L$ & $J$& $f^6$,\% & $jj$,\% & $\mu_{eff}$ & n$_f$ \\
\hline
     &      &2&.651 & .817 & .166 & 93.4 & 78.7 & .298 & \\
Pu1  & 3.27 &2&.518 & .656 & .138 & 94.5 & 83.1 & .283 & 5.71\\
     &      &2&.515 & .652 & .137 & 94.5 & 83.3 & .283 & \\
\hline
Pu2  & 4.63 &1&.356 & .449 & .094 & 96.3 & 88.5 & .243 & 5.70\\
\hline
\end {tabular}
\label {res_ES}
%\vspace {0.5cm}
\end {table}

\begin {figure}
\centerline {\includegraphics [clip=true,width=.85\columnwidth]
{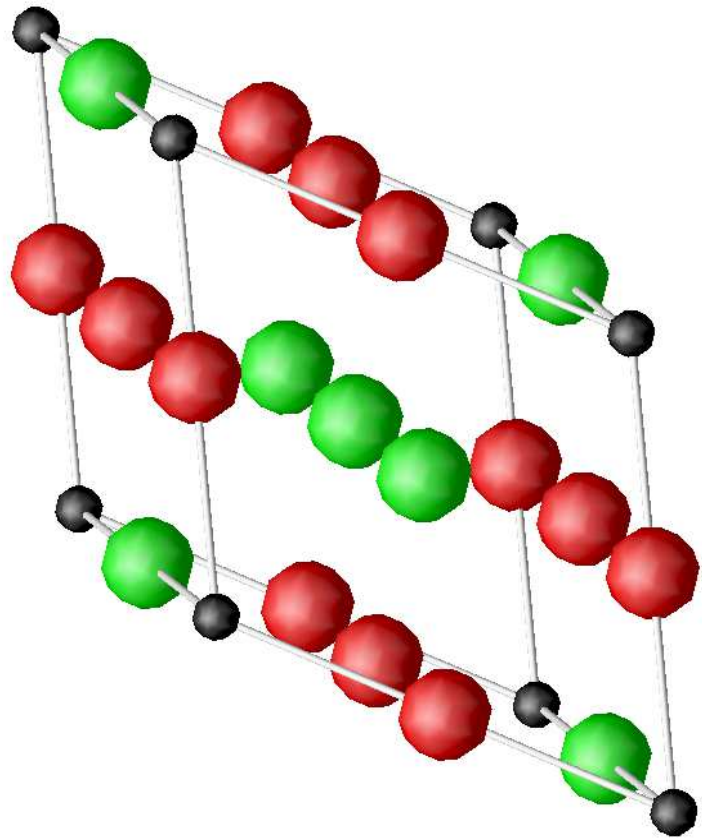} }
\caption {(Color online). Non-collinear order obtained for 8-atoms
supercell with vacancy. Vacancy is shown as black spheres. Green and red
spheres are Pu atoms with opposite directed total moments. The radii of
spheres are proportional to the magnitude of correspondent magnetic
moment.}
\label {afm-c.ES}
\end {figure}

Another type of defects appearing in fresh Pu during first several
years~\cite {Deremov} is vacancy. Small supercell consisting of 8 Pu
atoms was considered. One Pu atom was removed from its position in
the supercell and after relaxation of crystal structure this empty
space was artificially filled with empty sphere. Unfortunately, 8-atoms supercell is not sufficient to
describe correctly the relaxation of crystal structure within the
MEAM. Shifts of Pu atoms were obtained to be negligible. Nevertheless
 removing of one atom from the supercell lowers the
symmetry of crystal, since the local neighborhood of plutonium 
atoms becomes differnet. Note, that this turns out enough to give
rise to the local moments on Pu sites. Opposite to IS action, the
vacancy affects much larger on Pu1 that form the first
coordination sphere of it (Tab.~\ref {res_ES}). Magnetic moments of
Pu2 atoms that belong to the 2-nd coordination sphere are smaller.

The average value of the effective magnetic
moment in the supercell with vacancy is
$\sim$0.28~$\mu_B$. In contrast to the case of IS in 32-atoms supercell,
the resulted magnetic order is the analogue of C-type AFM (see Fig.~\ref
{afm-c.ES}).

These results prove that both types of defects call local magnetic
moment on Pu atoms due to distortion of fcc structure or even
lowering of symmetry. Different types of defects affect magnetism in
Pu in different ways: IS calls the largest magnetinc moment on an
atom at large distance whereas vacancy affects mostly its nearest
neighbors. Different types of defects result also in different types
of AFM order. Simultaneous effect of IS and vacancy could also give
rise to the local moments and lead to more complicated pattern of Pu
ions magnetic order.

\section {Vacancy and Interstitial Plutonium at minimal distance}
\label {ES_IS_close}

Since IS and vacancy affects on the magnetism in Pu in different ways, 
we can expect that their simultaneous influence could also give rise of local moment, 
and produce some complicated magnetic pattern.    

\begin {table}
\caption {Magnetic properties of Pu ions calculated for 32-atoms supercell
with one vacancy and IS at minimal distance. See also caption of Tab.~\ref
{res_IS_large}.}
%\vspace {0.5cm}
\centering
\begin {tabular}{c|c|c|c|c|c|c|c|c|c}
\hline
&D, \AA&n$_{atom}$& $S$ & $L$ & $J$& $f^6$,\% & $jj$,\% & $\mu_{eff}$ & n$_f$ \\
\hline
IS   &&1& 0.058 & 0.086 & 0.029 & 98.9 & 98.2 & 0.145 & 6.1 \\
\hline
\raisebox {-2ex}[1pt][1pt]{Pu1}  &\raisebox {-2ex}[1pt][1pt]{4.02}&2& 0.250 & 0.348 & 0.097 & 96.1 & 92.0 & 0.259 &
\raisebox {-2ex}[1pt][1pt]{5.61} \\
 &&2& 0.195 & 0.267 & 0.073 & 97.1 & 93.8 & 0.225 &      \\
\hline
Pu2  &2.70&1& 0.009 & 0.015 & 0.006 & 99.8 & 99.7 & 0.064 & 6.22 \\
\hline
\raisebox {-2ex}[1pt][1pt]{Pu3}  &\raisebox {-2ex}[1pt][1pt]{5.10}&2& 0.485 & 0.583 & 0.098 & 96.1 & 84.1 & 0.237 &
\raisebox {-2ex}[1pt][1pt]{5.71} \\
 &&2& 0.532 & 0.645 & 0.113 & 95.5 & 82.6 & 0.252 &      \\
\hline
\raisebox {-2ex}[1pt][1pt]{Pu4 } &\raisebox {-2ex}[1pt][1pt]{6.91}&2& 0.230 & 0.313 & 0.083 & 96.7 & 92.7 & 0.239 &
\raisebox {-2ex}[1pt][1pt]{5.75} \\
  &&2& 0.021 & 0.038 & 0.017 & 99.3 & 99.4 & 0.111 &      \\
\hline
\raisebox {-2ex}[1pt][1pt]{Pu5}  &\raisebox {-2ex}[1pt][1pt]{4.09}& 2&0.165 & 0.212 & 0.047 & 98.1 & 94.7 & 0.180 &
\raisebox {-2ex}[1pt][1pt]{5.71} \\
  &&2& 0.122 & 0.155 & 0.033 & 98.7 & 96.1 & 0.153 &      \\
\hline
\raisebox {-2ex}[1pt][1pt]{Pu6}  &\raisebox {-2ex}[1pt][1pt]{2.85}&2& 0.057 & 0.081 & 0.024 & 99.1 & 98.2 & 0.132 &
\raisebox {-2ex}[1pt][1pt]{5.78} \\
  &&2& 0.088 & 0.117 & 0.029 & 98.8 & 97.2 & 0.145 &      \\
\hline
Pu7  &5.18&2& 0.123 & 0.156 & 0.034 & 98.7 & 96.1 & 0.154 & 5.75 \\
\hline
Pu8  &6.91&1& 0.069 & 0.077 & 0.008 & 99.7 & 97.7 & 0.074 & 5.75 \\
\hline
Pu9  &5.35&2& 0.220 & 0.277 & 0.057 & 97.7 & 92.9 & 0.196 & 5.77 \\
\hline
Pu10 &6.99&2& 0.343 & 0.448 & 0.105 & 95.8 & 88.9 & 0.261 & 5.78 \\
\hline
\raisebox {-2ex}[1pt][1pt]{Pu11} &\raisebox {-2ex}[1pt][1pt]{5.24}&2& 0.069 & 0.095 & 0.027 & 98.9 & 97.8 & 0.140 &
\raisebox {-2ex}[1pt][1pt]{5.79} \\
 &&2& 0.150 & 0.199 & 0.049 & 98.0 & 95.2 & 0.186 &      \\
\hline
\end {tabular}
\label {res_ES_IS_close}
%\vspace {0.5cm}
\end {table}

\begin {figure} [b!]
\centerline {
\includegraphics [clip=true,width=.85\columnwidth]
{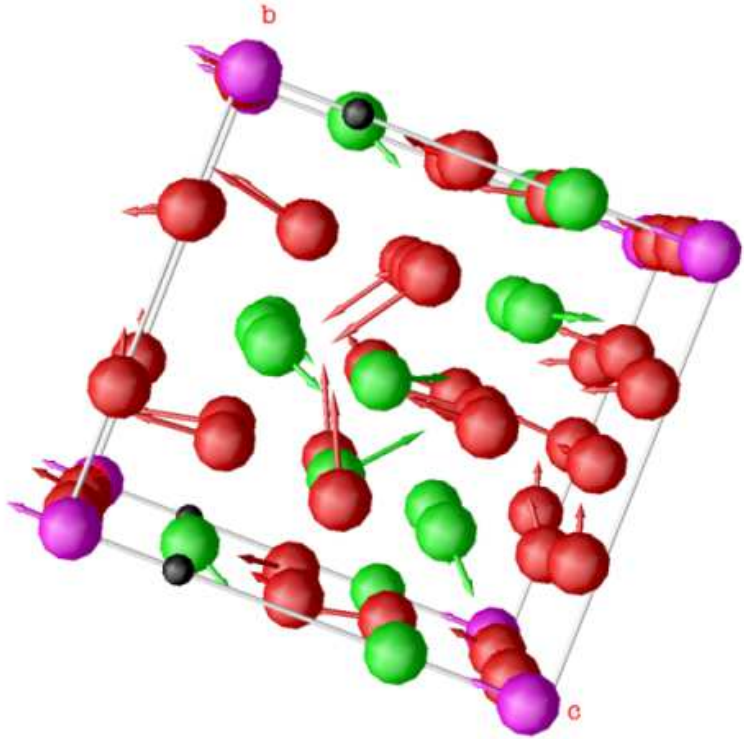} }
\caption {Ferrimagnetic order obtained for 32-atoms supercell with IS and
vacancy at minimal distance. Black spheres denote vacancy, purple spheres
-- IS. Green and red spheres are Pu atoms with opposite signs of
$z$-component of magnetic moment. The length of arrows is proportional to
the magnitude of correspondent effective magnetic moment.}
\label {ms_ES_IS_close}
\end {figure}

As the first step, 32-atoms supercell with IS and vacancy at minimal
distance has been investigated. Relaxation of the supercell was made with use of molecular dynamics
within MEAM. Both removing one Pu atom from its site and relaxation lowers the symmetry of considered
supercell and one calss of Pu has been devided in 11 classes. Six of them have 2 subclasses (see
Tab.~\ref{res_ES_IS_close}). Small local moments also developed
at Pu sites of relaxed supercell. Magnitude of moments and other
results of LDA+U+SO calculation are presented in Tab.~\ref
{res_ES_IS_close}. Two different values of moments for one type of
Pu atoms is the consequence of lowering of the symmetry of supercell
due to orbital polarization. Mutual action of IS and vacancy
decreases dispersion of magnitudes on differnet sites. The averaged
value of effecive moment at Pu atom is 0.18~$\mu_B$.

Three types of Pu atoms, Pu10, Pu1, and Pu3, have the largest
magnitudes of magnetic moment, 0.261~$\mu_B$, 0.259~$\mu_B$, and
0.252~$\mu_B$, correspondingly.  Atoms of type Pu1 and Pu3 are the
nearest to the vacancy (except IS) and have hence the largest
magnetic moments in agreement with results of our previous
calculation for vacancy in 8-atoms supercell (Sec.~\ref {ES_small}).
Pu6 atoms belong to the first coordination sphere of vacancy but
posses much smaller moment, $\sim$.14~$\mu_B$. This atom is
positioned in the first coordination sphere of IS and, in agreement
with our results for one IS in 32-atoms supercell, IS suppresses
magnetism on Pu6 atom.  Finally, Pu10 has a sizeable value of
magnetic moment but it is smaller then that at Pu atom in the center
of 32-atom supercell with single IS. This could be explained by the
action of vacancy that calls large local moments near itself and
suppresses the magnetism on far standing atoms.

Simultaneous effect of IS and vacancy results in more complicated canted
AFM pattern, which could not be identified with any standart type.
Calculated canted AFM order is presented in Fig.~\ref {ms_ES_IS_close}.

\section {Vacancy and Interstitial Plutonium at large distance}

\begin {table}
\caption {Magnetic properties calculated for 32-atoms supercell with one
vacancy and IS at large distance. See also caption of Tab.~\ref
{res_IS_large}.}
%\vspace {0.5cm}
\centering
\begin {tabular}{c|c|c|c|c|c|c|c|c|c}
\hline
&D,\AA& n$_{atoms}$&$S$ & $L$ & $J$& $f^6$,\% & $jj$,\% & $\mu_{eff}$ & n$_f$ \\
\hline
IS   &    &1& .052 & .083 & .032 & 98.7 & 98.4 & 0.153 & 6.12\\
Pu1  &4.26&4& .144 & .176 & .032 & 98.7 & 95.4 & 0.149 & 5.70\\
Pu2  &2.79&2& .103 & .141 & .038 & 98.5 & 96.7 & 0.164 & 5.78\\
Pu3  &5.01&1& .235 & .333 & .098 & 96.0 & 92.6 & 0.261 & 5.73\\
Pu4  &6.67&2& .152 & .188 & .035 & 98.6 & 95.1 & 0.157 & 5.72\\
Pu5  &3.77&4& .268 & .361 & .093 & 96.3 & 91.4 & 0.250 & 5.70\\
Pu6  &3.11&1& .065 & .091 & .025 & 99.0 & 97.9 & 0.136 & 5.73\\
Pu7  &5.25&2& .031 & .039 & .007 & 99.7 & 99.0 & 0.072 & 5.74\\
Pu8  &7.12&1& .173 & .219 & .046 & 98.2 & 94.4 & 0.178 & 5.71\\
Pu9  &2.63&1& .077 & .128 & .051 & 98.0 & 97.7 & 0.194 & 6.01\\
Pu10 &4.90&2& .226 & .276 & .050 & 98.0 & 92.7 & 0.182 & 5.75\\
Pu11 &6.83&1& .169 & .276 & .107 & 95.7 & 94.8 & 0.280 & 5.73\\
Pu12 &2.79&2& .020 & .031 & .012 & 99.5 & 99.4 & 0.093 & 5.82\\
Pu13 &4.87&2& .139 & .181 & .042 & 98.3 & 95.5 & 0.172 & 5.78\\
Pu14 &7.29&2& .073 & .131 & .058 & 97.7 & 97.8 & 0.208 & 5.75\\
Pu15 &5.23&2& .239 & .303 & .065 & 97.4 & 92.3 & 0.208 & 5.77\\
Pu16 &5.42&1& .164 & .228 & .064 & 97.5 & 94.8 & 0.212 & 5.78\\
Pu17 &5.02&1& .245 & .312 & .068 & 97.3 & 92.1 & 0.213 & 5.79\\
\hline
\end {tabular}
\label {res_ES_IS_far}
%\vspace {0.5cm}
\end {table}

Finally, we made the same calculation for the 32-atom supercell contained IS and
vacancy at the large distance. As well as in previous cases relaxation of crystal structure has been
made within MEAM before band structure calculation. Considered defects also lower the symmetry and 17
new Pu classes arouse. The values of moments and contribution of
coupling types and electronic configuration are presented in Tab.~\ref
{res_ES_IS_far}.

\begin {figure}[t!]
\centerline {
\includegraphics [clip=true,width=.85\columnwidth] {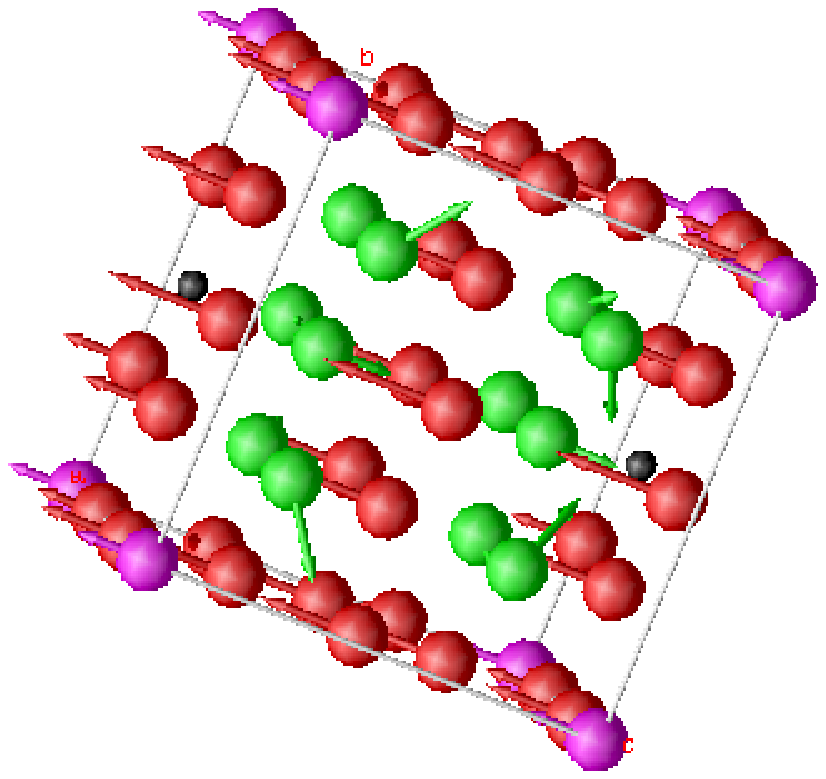} }
\caption  {Ferrimagnetic order obtained for 32-atoms supercell with IS and
vacancy at large distance.  See also caption of Fig.~\ref {ms_ES_IS_close}.
}
\label {ms_ES_IS_far}
\end {figure}

Like our calculation for IS and vacancy at minimal distance, local
moments developed at all Pu atoms. The magnitude of moments depends on the
distances to both IS and vacancy. Incommensurate magnetic order with strong
noncollinearity was obtained for this type of defects position (see
Fig.~\ref{ms_ES_IS_far}). The mechanism of the formation of magnetic
moments was described above. The averaged magnetic moment on Pu is
0.179~$\mu_B$.

\section {Conclusions}

Band structure calculations have been run for 4
supercells contained IS, vacancy, and both IS and vacancy at small
and large distances. For the supercell with one IS ferrimagnetic
order close to A-type AFM was obtained. The maginudes of local
moments are  0.06--0.46~$\mu_B$ and the averaged moment is
0.26~$\mu_B$. The largest value of magnetic moment has atoms at the
largest distance from IS.  Ferrimagnetic order close to C-type AFM
was obtained for 8-atoms supercell with vacancy. This type of defect
calls the largest moment on Pu atoms in first coordination sphere.
Magnetic moments obtained in 32-atoms supercell with both IS and
vacancy have smaller dispersion of magnitude 0.1--0.3~$\mu_B$ and
smaller averaged moment $\sim$0.18~$\mu_B$. Simultaneous action of
these defects results in incommensurate magnetic order with strong
non-collinearity. 
Nevertheless the long range order obtained 
in the present work should be considered as an artifact of computation method. 
Our results indicate that sort range order could appear due to defects in fcc Pu. 
And the type of such order depends strongly on the distance to the corresponding defect.
The implementation of the LDA+DMFT method is necessary to describe  more 
accurately the magnitudes of 
local moment in paramagnetic phase of Pu. 

Results of calculation explain presence of magnetic moment in aged Pu
samples and agree well with experimental data~\cite {Verkhovskii05,
Verkhovskii07, Fluss06}.

Support by the Russian Foundation for Basic Research under Grant No.
RFBR-07-02-00041 is gratefully acknowledged.

\end {document}